\documentclass{aastex62}

\usepackage{soul}

\usepackage{lineno}
\usepackage{hyperref}

\graphicspath{{./}{figures/}}

\received{December 6, 2019}
\revised{\today}
\submitjournal{The Astrophysical Journal}

\shorttitle{A Dust Trail in Phaethon's Orbit}
\shortauthors{Battams et al.}
\begin{document}

\title{Parker Solar Probe Observations of a Dust Trail in the Orbit of (3200) Phaethon}

\correspondingauthor{Karl Battams}
\email{karl.battams@nrl.navy.mil}

\author[0000-0002-8692-6925]{Karl Battams}
\affil{US Naval Research Laboratory, 4555 Overlook Avenue, SW, Washington, DC 20375, USA}

\author[0000-0003-2781-6897]{Matthew M. Knight}
\affiliation{University of Maryland, College Park, Maryland, 20742 USA}

\author[0000-0002-6702-7676]{Michael S.P. Kelley}
\affiliation{University of Maryland, College Park, Maryland, 20742 USA}

\author[0000-0002-8353-5865]{Brendan M. Gallagher}
\affil{US Naval Research Laboratory, 4555 Overlook Avenue, SW, Washington, DC 20375, USA}

\author[0000-0001-9027-8249]{Russell A. Howard}
\affil{US Naval Research Laboratory, 4555 Overlook Avenue, SW, Washington, DC 20375, USA}

\author[0000-0001-8480-947X]{Guillermo Stenborg}
\affil{US Naval Research Laboratory, 4555 Overlook Avenue, SW, Washington, DC 20375, USA}



\begin{abstract}

We present the identification and preliminary analysis of a dust trail following the orbit of (3200) Phaethon as seen in white light images recorded by the Wide-field Imager for Parker Solar Probe (WISPR) instrument on the NASA Parker Solar Probe (\textit{PSP}) mission. During \textit{PSP}'s first solar encounter in November 2018, a dust trail following Phaethon's orbit was visible for several days and crossing two fields of view. Preliminary analyses indicate this trail to have a visual magnitude of 15.8 $\pm$0.3 per pixel and a surface brightness of 25.0 mag arcsec$^{-2}$ as seen by \textit{PSP}/WISPR from a distance of $\sim$0.2~au from the trail. We estimate the total mass of the stream to be $\sim(0.4-1.3){\times}10^{12}$~kg, which is consistent with, though slightly underestimates, the assumed mass of the Geminid stream but is far larger than the current dust production of Phaethon could support. Our results imply that we are observing a natural clustering of at least some portion of the Geminid meteor stream through its perihelion, as opposed to dust produced more recently from perihelion activity of Phaethon.

\end{abstract}


\section{Introduction} \label{s:intro}
Discovered in 1983 \citep{IAUC3878}, asteroid (3200) Phaethon is one of the most widely-studied inner solar system minor bodies, by virtue of a 1.434 year orbit, its large size for a near-Earth object \citep[6~km in diameter,][]{taylor19}, and a low 0.0196~au Earth minimum orbit intersection distance (MOID) favorable to ground-based optical and radar observations. Phaethon is recognized as the parent of the Geminid meteor shower and is associated with the Phaethon-Geminid meteoroid stream complex that includes likely relationships with asteroids 2005 UD and 1999 YC \citep[e.g.,][]{IAUC3881,Gustafson1989,Williams1993,Ohtsuka2006,Ohtsuka2008}. The link between Phaethon and the Geminids is based upon the similarity of their orbits and dynamical integrations; despite repeated efforts, no cometary activity of Phaethon has ever been detected from the ground \citep[e.g.,][]{Cochran1984,Chamberlain1996,Hsieh2005,Wiegert2008}. Phaethon is now generally considered an ``activated'' asteroid \citep{Jewitt2012} rather than an inactive comet due to numerous asteroid-like properties including Tisserand parameter (4.510, JPL Horizons), albedo \citep[0.11,][]{Green1985}, and spectral shape \citep[B-type, e.g.,][]{Licandro2007}. Furthermore, \citet{Deleon2010} identified a dynamical pathway showing it could plausibly be an escaped member of the Pallas family from the main asteroid belt.

Due to a small perihelion distance of just 0.14 au, observations of Phaethon near the Sun are impossible from traditional ground-based telescopes. NASA's {\it Solar and Heliospheric Observatory} \citep[{\it SOHO};][]{Brueckner1995} has observed the near-Sun region out to 0.15~au continuously since 1996 but, despite detecting more than 3000 other small bodies \citep{Battams2017}, has never detected Phaethon. The first detections of Phaethon at perihelion were made by the Sun-Earth Connection Coronal and Heliospheric Investigation \citep[SECCHI,][]{Howard2008} on NASA's {\it Solar Terrestrial Relations Observatory} \citep[{\it STEREO};][]{Kaiser07} which launched in 2006. During favorable apparitions in 2009, 2012, and 2016, observations obtained by the SECCHI Heliospheric Imager-1 \citep[HI-1,][]{Eyles2008} instrument revealed unexpected brightening and the presence of a short-lived tail $\sim$0.1$^\circ$ in length \citep{Jewitt2010,Jewitt2013,Li2013,Hui2017}. These observations unequivocally demonstrated that Phaethon is active near perihelion, but found the mass-loss rate orders of magnitude too low to sustain the Geminids.

In 2014, a re-examination of infrared observations from the Diffuse Infrared Background Experiment (DIRBE) on the {\it Cosmic Background Explorer (COBE)} revealed a dust trail in the orbit of Phaethon \citep{Arendt2014}. Analyses of these observations were largely qualitative rather than quantitative, but notable results were that the trail was detected far from Phaethon itself and was described to ``brighten dramatically'' as the Earth passed through Phaethon's orbital plane. Searches for an optical trail both near perihelion \citep{Hui2017} and at larger heliocentric distances \citep{Urakawa2002,Ishiguro2009} have been unsuccessful. Phaethon's extraordinary close approach to Earth in 2017 December (geocentric distance $<$0.07 au) allowed very high resolution searches for large dust grains \citep{Jewitt2019} and small fragments \citep{Jewitt2018,Ye2018}, but none were detected, further supporting the conclusion that the Geminids are not in a steady state.

In this article we present and analyze detections of a white light dust trail following the orbit of Phaethon, seen by the Wide-field Imager for Parker Solar Probe (\textit{WISPR}) instrument on the NASA Parker Solar Probe (\textit{PSP}) mission. In $\S$\ref{s:obs} of this paper, we describe the WISPR observations, summarize the data processing steps, and present a selection of images from the study. $\S$\ref{s:analysis} presents an analysis of the properties of this dust trail, including the photometric properties of the trail and estimates of the mass. In $\S$\ref{s:discussion} we provide some discussion of the results of this study and outline some of the next steps for future works. We also demonstrate and discuss the overlap of the Geminid stream with the Phaethon orbit as observed from \textit{PSP}'s location during the encounter. Finally, $\S$\ref{s:conclusions} provides a brief summary of the conclusions of this study.

\section{Observations}
\label{s:obs}
\begin{figure}[ht!]
\plotone{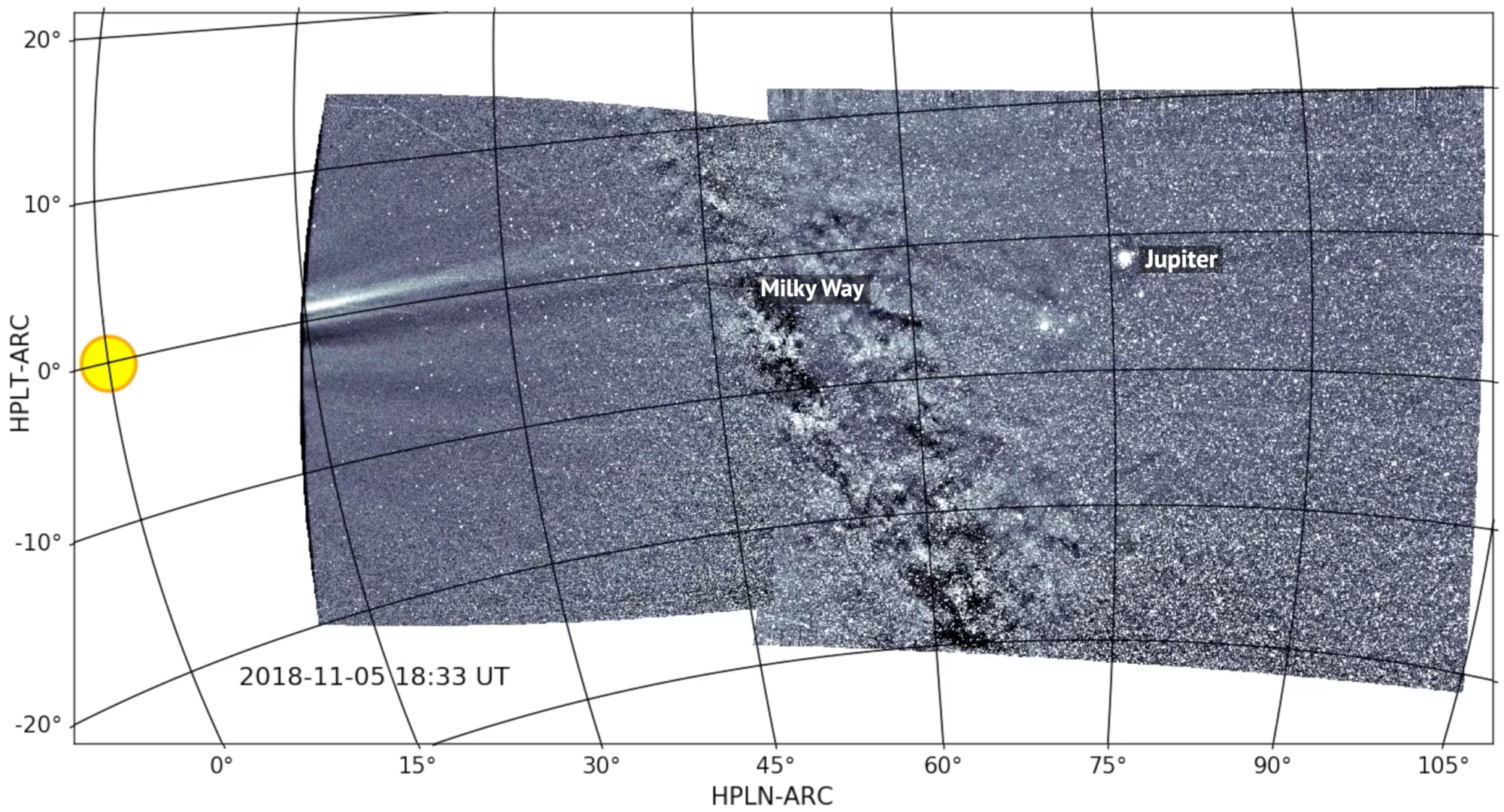}
\caption{Composite image showing the inner (left) and outer (right) WISPR fields of view, with a pair of observations recorded 2018 November 05 16:12~UT. A helio-projective longitude and latitude grid is superimposed upon the image. The yellow circle indicates the true location of the Sun (to scale), $13.5^{\circ}$ to the left and outside of the WISPR field of view, enabling the instrument to view into the direction of spacecraft motion, to the west of the Sun. The center of the field of view is about $2.5^{\circ}$ to the south of the heliographic equator. The Milky Way and planet Jupiter are labelled.
\label{fig:wispr-widefield}}
\end{figure}

{\it PSP} \citep{Fox15} launched in August 2018 on an evolving seven-year, 24-perihelion orbit that provides the spacecraft with successively close encounters with the Sun from 35 solar radii (R$_\odot$, where 1~R$_{\odot} = 695,000~\mathrm{km} = 0.00465$~au) on its first solar encounter to less than 10~R$_\odot$ on the final planned encounters. WISPR \citep{Vourlidas15} is the only imaging instrument aboard {\it PSP} and comprises an overlapping pair of broadband, white-light heliospheric imagers designed to observe coronal structures and outflows over an evolving field of view. The specifications of these imagers, which use advanced pixel sensor (APS) detectors and are known as ``Inner'' (WISPR-I) and ``Outer'' (WISPR-O), are given in Table~\ref{tab:wispr-specs}, and an example image is shown in Figure \ref{fig:wispr-widefield}.

\begin{table}[ht!]
\centering
\begin{tabular}{ |p{5cm}||p{2.5cm}|p{2.7cm}|  }
    \hline
    \multicolumn{3}{|c|}{WISPR Telescopes} \\
    \hline
    Attribute & Inner (WISPR-I) & Outer (WISPR-O)\\
    \hline
    Focal Length ($f$) [mm] & 28 & 19.8\\
    Aperture [mm$^2$] & 42 & 51\\
    Bandpass [nm] & 490 - 740 & 475 - 725\\
    APS Resolution [pixels] & 1920$\times$2048 & 1920$\times$2048\\
    APS Plate Scale [arcmin/pixel] & 1.2 & 1.7\\
    Field of View [$^{\circ}$elongation] & 13.5$^{\circ}$ - 53$^{\circ}$ & 50$^{\circ}$ - 108$^{\circ}$\\
    \hline
\end{tabular}
\caption{Specifications of the WISPR cameras. \label{tab:wispr-specs}}
\end{table}

Observations from WISPR are only recorded during so-called ``encounter periods'', with the first such encounter spanning two weeks from 2018 October 31 through 2018 November 11. In this article we focus on a subset of observations recorded from the WISPR-I camera during \textit{PSP}'s first encounter. A second set of observations from the second encounter (2019 March 30 through 2019 April 10) are also available but their analysis is not presented here. However we note that the observations from the second encounter are essentially identical to those in the first encounter.

During the first encounter, WISPR began recording observations on 2018 November 01 at 00:00~UT, when the spacecraft was at 0.239~au from the Sun. \textit{PSP}'s perihelion occurred on November 06 at 03:28~UT, with the spacecraft reaching 0.166~au and a peak velocity of 95.328 kms$^{-1}$ relative to the Sun. Synoptic observations ceased on 2018 November 10 at 17:29~UT, with the spacecraft at 0.227 au. During the encounter, WISPR-I recorded 546 images at resolutions of 960$\times$1024 pixels and 23 images at 1920$\times$2048 pixels. Each observation was a stack of eight 2.56-second exposures (for a total exposure of 20.48 seconds), summed on board, and with a 20.36-second gap between each individual exposure. The data rate varied as a function of heliocentric distance. During the `outer perihelion', the first and last 2 days of the encounter, observations were recorded every 45 minutes. During the `inner perihelion', the 40 hours surrounding perihelion, observations were recorded every 7.8 minutes. In the `middle perihelion', the time between the `outer perihelion' and `inner perihelion', observations were recorded every 33.6 minutes. The effective field of view for WISPR-I ranged from elongations of 12.1--47.5 R$_\odot$ at the beginning of the encounter, to 8.4--33.0 R$_\odot$ at perihelion.

WISPR observations are available as Level-1 (raw, L1), Level-2 (L2), and Level-3 (L3), and can be obtained via the WISPR project website, \url{https://wispr.nrl.navy.mil/wisprdata}. The L2 data are the primary data product, calibrated to units of mean solar brightness and corrected for bias, vignetting and distortion. Due to the current availability of only two relatively short data collection sets in the two encounters, and inconsistencies between the types of observations recorded in the encounters thus far, calibration of the data is not straight-forward. Preliminary photometric calibrations, provided by means of analyzing stellar transits across the observations, have been completed using the first two PSP encounters, and appear self-consistent, though unquantified small uncertainties may still exist while the response of the APS detectors is still being fully characterized (\textit{Hess, Private Communication, September 2019}). Photometric calibration is discussed further in $\S$\ref{s:photom}.

Processing of the WISPR observations from L2 to a background-removed L3 product is challenging due primarily to the dominant signal in the raw data being that of the F-corona - sunlight scattered by the dust particles in orbit around the Sun \citep{Howard2019}. A commonly used approach to reveal the faint K-corona (electron) signal in coronagraph and heliospheric images acquired near 1 au, is to remove the F-corona excess \citep[e.g.,][]{Morrill2006} by performing a simple `background subtraction'. This process generally involves creating an average (or median) background image based on a several-weeks span of images, which can then be subtracted from the data to reveal the fainter K-corona structures and solar outflow. However, the rapidly evolving WISPR field of view does not allow for such an algorithm to be employed \citep[see, e.g.][]{stenborg17}. Instead, an F-corona model must be derived for each individual calibrated observation prior to being subtracted from the data to reveal the faint structures of interest. \cite{stenborg17} developed a state-of-the-art technique to prove this concept on {\it STEREO} SECCHI HI-1 images in preparation for WISPR. The full details of the algorithm as adapted to WISPR imagery are beyond the scope of this article, and will soon appear in \cite{stenborg20}. Briefly, for WISPR the technique exploits the smoothness of the background intensity (which is dominated by the F-corona) in the North-South direction to separate it from the K-corona's discrete features. There does exist a trade-off with the photometric accuracy of the resulting background-removed product, though in testing we again found the photometric accuracy of the Level-3 data product consistent with that of Level-2, but absent the dominant stray/scattered light signal that overwhelms the data.

This L3 data reveals a complex scenery in the WISPR images. An example of an L3 image is shown in Figure~\ref{fig:wispr-widefield}, which presents a composite of both the WISPR-I and -O fields of view recorded on 2018 November 05 16:12~UT with a helio-projective cartesian (HPC) longitude and latitude grid superimposed upon the image (black grid lines). %The images are displayed using a zenithal equidistant (ARC) projection \citep{Greisen1983}.
In the Figure, K-corona features (the streamer at about $0^\circ$ latitude) are revealed along with the star field, Jupiter, and the Milky Way. The Sun (yellow circle, shown to scale from \textit{PSP}'s location) always remains outside of the WISPR-I field of view as indicated in this figure. The orbital characteristics of the mission causes the Milky Way to cross the field of view of both detectors over several days in every encounter, making the computation of the background intensity more difficult.

\begin{figure}[ht!]
\includegraphics[scale=0.36]{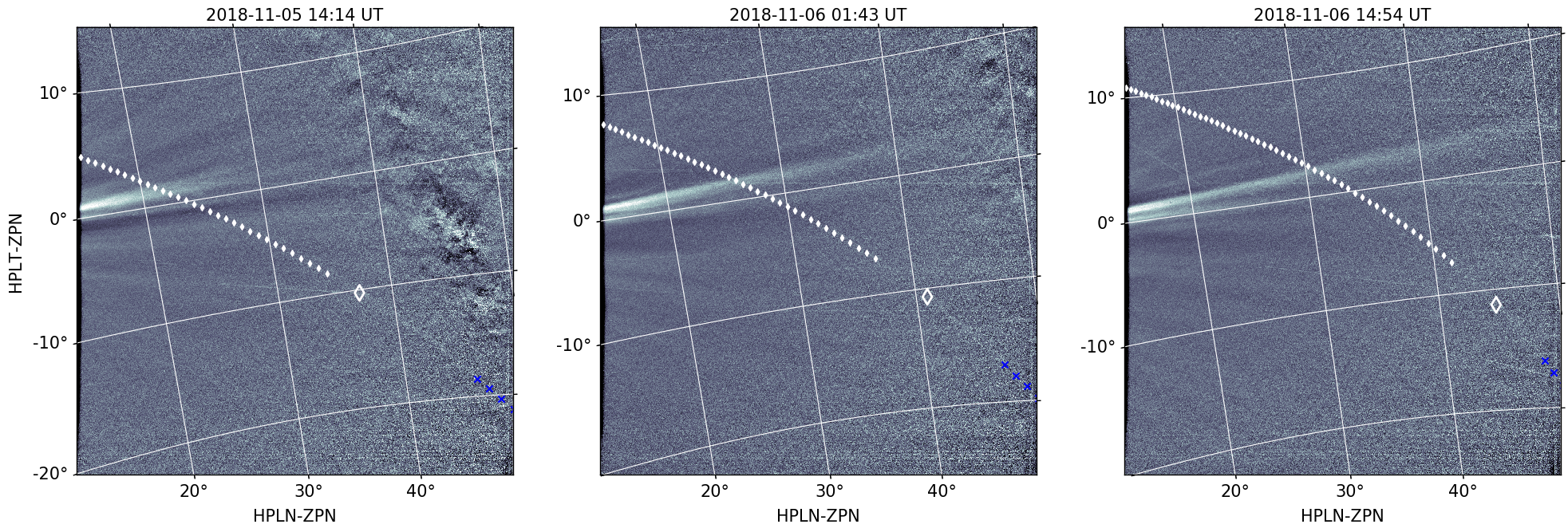}
\caption{Three WISPR-I observations recorded on November 5 14:14~UT, November 6 1:43~UT and November 6 14:54~UT, with a helio-projective longitude/latitude grid superimposed upon the data. Symbols are plotted along the path of Phaethon's orbit indicating the imaginary position of Phaethon along that orbit in 60 minute increments, with blue crosses indicating pre-perihelion and white dots post-perihelion. The symbols are excluded in the region of the data where the trail is most easily visible. The white diamond indicates the perihelion position of the orbit in the field of view.
\label{fig:wispr-threepanel}}
\end{figure}

An unexpected white-light feature revealed in the L3 data was the presence of a faint extended trail seen most clearly during the time period between approximately 2018 November 5 12:00~UT and 2018 November 6 12:00~UT, though continuing to be just barely visible in animated sequences of observations up until the end of the encounter on November 11. In Figure~\ref{fig:wispr-threepanel} we show three observations, recorded on November 05 14:14~UT, November 06 1:43~UT and November 06 14:54~UT, that highlight the most visible portion of this dust trail, which we see following perfectly along the orbital path of Phaethon. These images are created from the L3 data product but, to enhance the visibility of the signal, have been processed with a point (sigma) filter to reduce the presence of stars and sporadic dust streaks. (This process destroys photometry and thus was not applied to data used for quantitative analyses.) Again a heliocentric longitude/latitude grid has been superimposed upon the data using a zenithal polynomial projection (ZPN), which is a generalization of the ARC projection that adds polynomial terms in the zenith direction \citep{Greisen1983}. At the time these data were recorded, Phaethon was at a distance of approximately 2.39~au from the Sun, three weeks after its aphelion (2018 October 14; 2.403 au).

\begin{figure}[ht!]
\centering
\includegraphics[scale=0.36]{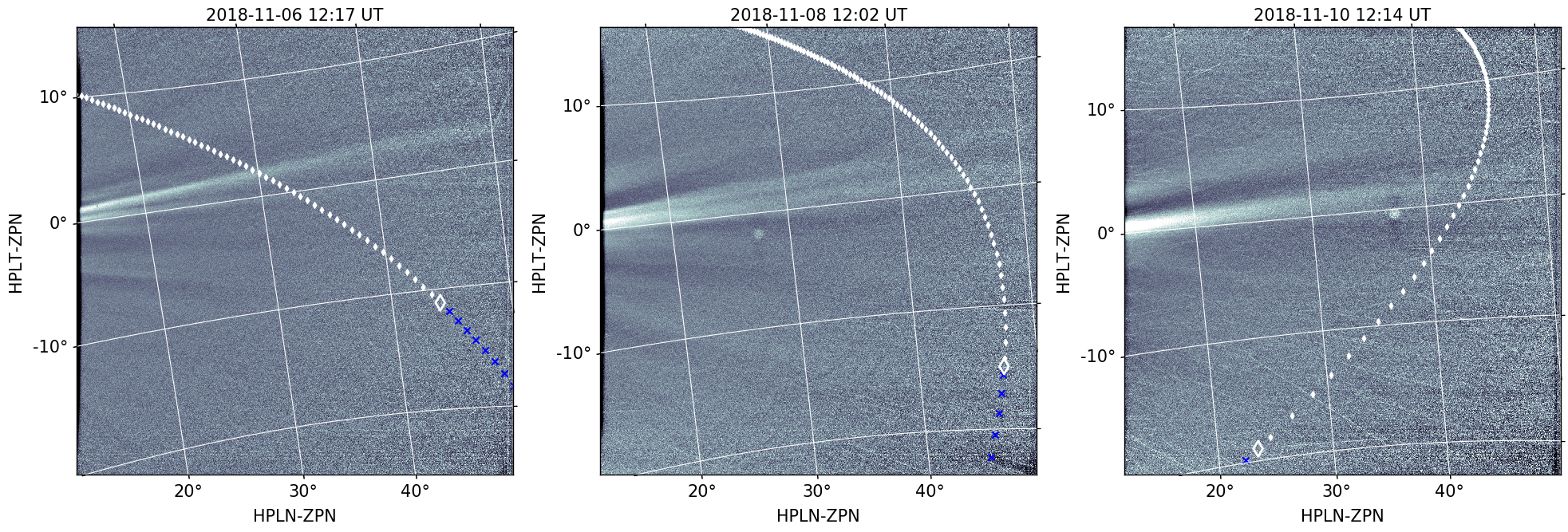}

\caption{A series of three WISPR observations illustrating the changing viewing geometry of the Phaethon orbit throughout the \textit{PSP} encounter. Again, symbols are plotted along the path of Phaethon's orbit indicating the imaginary position of Phaethon along that orbit in 60 minute increments, with blue crosses indicating pre-perihelion and white dots post-perihelion locations, and a white diamond indicating the perihelion point.
\label{fig:trail-evolution}}
\end{figure}

Our analyses in this article focus primarily on the dust trail seen in the WISPR-I FOV observations recorded 2018 November 5--6. As noted, the trail was visible in observations outside of this range but at a detection level too low for reliable interpretation. We also exclude the small portion of the path that crosses the WISPR-O field of view as a number of photometric and calibration uncertainties and non-linearities in these observations currently preclude reliable analysis, though will be considered in future studies of the data set.

Figure~\ref{fig:trail-evolution} shows the evolution of the viewing geometry of the dust trail throughout \textit{PSP}'s first encounter, with observations taken 2018 November 06 12:17~UT, 2018 November 08 12:02~UT and 2018 November 10 12:14~UT. Again these images are based upon the L3 data product with a filter applied to minimize the brightness of stars. A white diamond again indicates the location of Phaethon's perihelion in each panel of the figure. The trail was only barely visibly detectable in data beyond 2018 November 06, and generally only in animated sequences of the observations. Thus the trail is not visually detectable in the still frames shown in the second and third panels of this Figure. Also, for this same reason, we only present analyses of the data recorded during the early part of the encounter.

\begin{figure}[ht!]
\centering
\hspace*{+1.2cm}\includegraphics[scale=0.6]{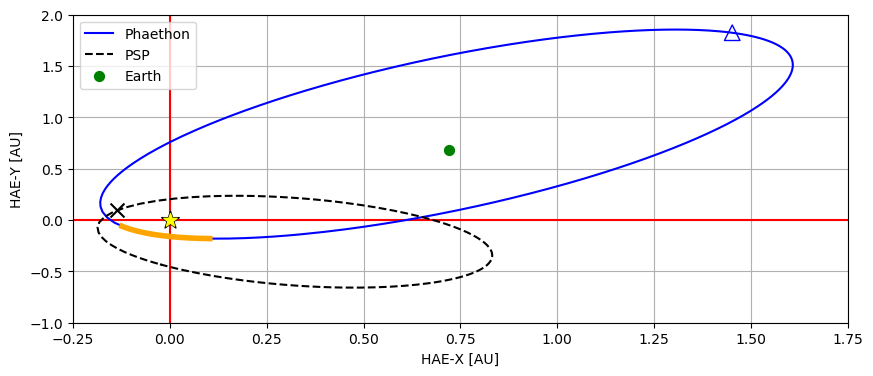}
\newline
\includegraphics[scale=0.60]{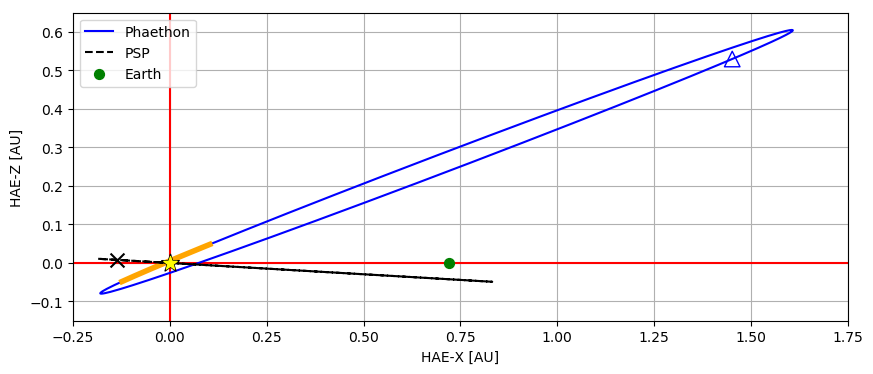}%
\caption{Schematic of the orbit of Phaethon (blue) and {\it PSP} (dashed black) as seen from above the solar system (upper panel) and from the side (lower panel) on November 06, 2018 at 1:43UT, corresponding to the image shown in Figure~\ref{fig:wispr-threepanel}. The blue triangle indicates the location of Phaethon at the time of observation, the green circle the location of Earth, the yellow star the Sun location and the black X the location of \textit{PSP}. The thick orange arc along the Phaethon orbit indicates the portion of the orbit crossing the WISPR field of view at this time, and the bold red axes highlight the XY (top) and XZ (lower) axes.
\label{fig:orbit-diag}}
\end{figure}

Figure~\ref{fig:orbit-diag} provides a schematic of the orbits of Phaethon (blue) and \textit{PSP} (dashed black) as observed on November 06, 2018 at 1:43~UT, corresponding to the middle panel of the image shown in Figure~\ref{fig:wispr-threepanel}. The thick orange arc along the Phaethon orbit indicates the portion of the orbit contained within the WISPR field of view at this time, and corresponds to a physical length of approximately 42-million km, or $\sim$5\% of Phaethon's $\sim$6~au orbital path around the Sun.

Our analysis of these observations, presented in the following Section, focus on the basic observational properties of the trail ($\S$\ref{s:obs-properties}), the photometric properties ($\S$\ref{s:photom}), and the dust properties ($\S$\ref{s:dust-properties}).

\section{Analysis} \label{s:analysis}

\subsection{Basic observational properties\label{s:obs-properties}}

\begin{figure}[ht!]
\centering
\includegraphics[scale=0.3]{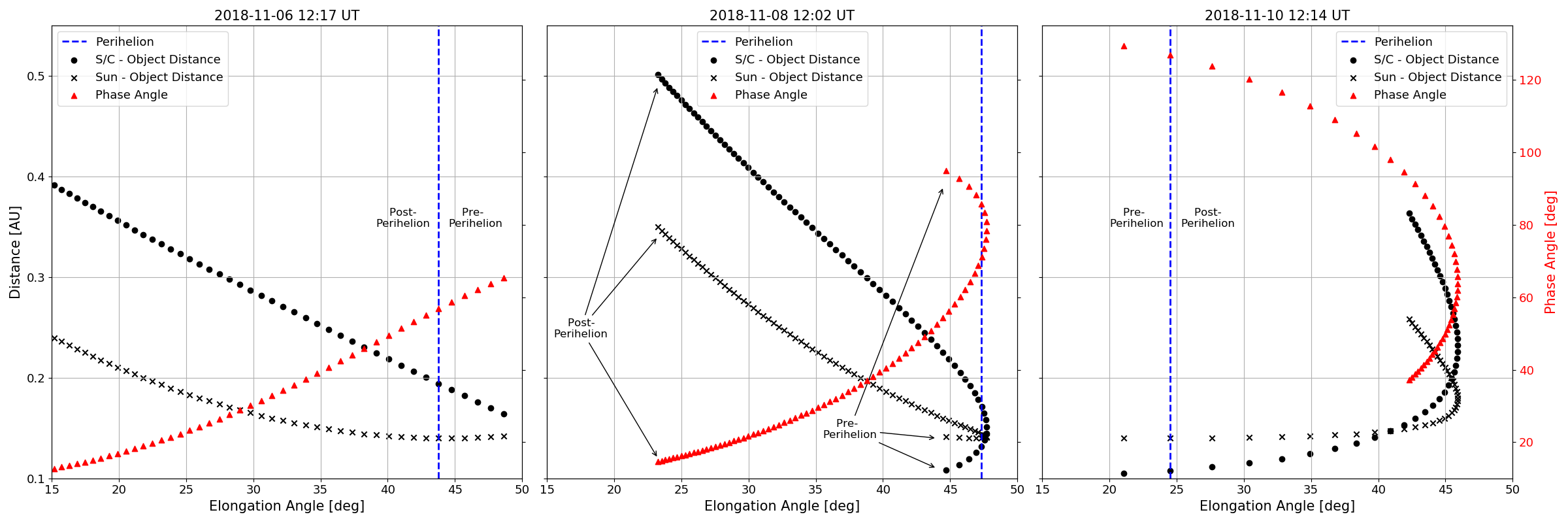}
\caption{The distance between the spacecraft (S/C) and Phaethon's orbit (black circles), the distance between the Sun and Phaethon's orbit (black crosses), and the phase
angle of Phaethon's orbit as viewed by PSP (red triangles), as a function of elongation angle at dates corresponding to the three panels of Figure~\ref{fig:trail-evolution}. The elongation angle shown here corresponds directly to HPLN-ARC shown in Figure~\ref{fig:wispr-widefield} and the phase angle (red triangles) represents the phase angle of Phaethon's orbit as viewed from PSP. Pre- and post-perihelion data points are indicated by the vertical dashed line (and labels/arrows in the middle panel).
\label{fig:dist-phase-angle-plot}}
\end{figure}

The dust trail was not visually detected across the entire field of view, though was visible across much of the field of view on 2018 November 05 - 06, and detectable by-eye in animations spanning most of the encounter. Throughout the encounter, the phase angles along the trail and the trail-spacecraft distance varied significantly as a consequence of \textit{PSP}'s very rapid passage through the inner heliosphere. In Figure~\ref{fig:dist-phase-angle-plot} we plot the distance between the spacecraft and Phaethon's orbit (black circles), the distance between the Sun and Phaethon's orbit (black crosses), and Phaethon's phase angle (red circles), as a function of elongation angle at dates corresponding to the three panels of Figure~\ref{fig:wispr-threepanel}. All three panels share the same \textit{y-axis} to enable easier comparison of both distance (left-hand y-axis) and phase angle (right-hand y-axis) over time.

As we are observing approximately the same section of the Phaethon orbit throughout the entire encounter, the Sun-Trail (``Sun-Object'') distances remain largely constant. The Spacecraft-Trail (``S/C-Object'') distances, however, vary quite significantly. At the beginning of the encounter (November 05--06, first panel of Figure~\ref{fig:wispr-threepanel}), \textit{PSP} was approximately 0.17 -- 0.4~au from the trail, with the most visible portion of the trail being that closest to the spacecraft. During the middle of the encounter ($\sim$November 08, second panel of Figure~\ref{fig:wispr-threepanel}), more distant parts of the trail - up to 0.5~au from \textit{PSP} - are entering the field of view, though a small portion of the trail at the largest elongation angles is as close as 0.11~au from \textit{PSP}. The trail is extremely difficult to visually detect during these dates. Towards the end of the encounter ($\sim$November 10, third panel of Figure~\ref{fig:wispr-threepanel}), the majority of the trail that crosses the WISPR-I field of view is very close to the spacecraft ($\sim$0.1 -- 0.2~au), with just a small portion up to 0.3~au from \textit{PSP}. The trail is again very challenging to detect during this period, though is slightly more visible towards the end of the encounter than during the middle. Throughout the encounter, the trail appears centered almost perfectly along the orbit of Phaethon - a point that becomes relevant in later discussion.

Assessing the role that phase angle plays in the observability of the trail is not trivial. The red triangles in Figure~\ref{fig:dist-phase-angle-plot} represent the phase angle of Phaethon's orbit as viewed from PSP. We used these to evaluate the phase function everywhere along the trail using the {\it HG} parameters for Phaethon given in \citet{tabeshian19}. Here we are, in essence, assuming the dust we are observing to be ``mini-Phaethons'' with a behavior different than for typical comet dust. Absent a detailed modeling effort beyond the scope of this article, it is difficult to fully interpret the trail visibility in the context of the phase angles presented in this Figure. Broadly speaking, however, we see that for much of the encounter (at least $\sim$November 05 -- 08), the phase angles evolve relatively slowly and span the $\sim$10--75$^\circ$  range. For the last two or three days there is a noticeable shift away from small phase angles and small elongations to much larger - over 120$^\circ$ - phase angles at small elongations. The most visually detectable portions of the trail appear to be those between 40--60$^\circ$, although this is only a broad generalization and not a universal rule throughout this (or the second) \textit{PSP} encounter.

\subsection{Photometric Properties} \label{s:photom}

Extracting and isolating the signal of the dust trail from the WISPR observations is extremely challenging, with determination of the photometric brightness of the trail hampered by several factors. First, the field of view contains many stars, the signal from which tremendously outweigh the signal we are seeking to isolate, and securing a suitable set of aperture backgrounds of any reasonable size without the inclusion of sporadic stars, was essentially impossible. Second, and to compound the first problem, many of the WISPR observations contain large dust particle events \citep{stenborg20} which also destroy attempts to secure a stable background\footnote{An example of such an image can be found at the WISPR project website: \url{https://wispr.nrl.navy.mil/data/rel/pngs/L3/orbit02/inner/20190402/psp\_L3\_wispr\_20190402T042924_V1_1221.png}}. A third complication is that the spacecraft was traveling on the order of 80~km~s$^{-1}$ at the time of observation, and thus the viewing geometry, field of view, and illumination of solar system dust at rapidly changing heliocentric distances, were evolving continuously. The spacecraft motion makes the process of selecting suitable aperture backgrounds particularly challenging as we must first select regions in the observations based on their HPC location, not their pixel location, and secondly correct the HPC coordinates by a scale factor to account for the variable heliocentric distance. The latter steps are essential to ensure that the region of space we are observing in any given frames remains approximately the same. In the case of observations that are temporally close to one another (less than $\sim3-4~hours$) such as were used in this study, the latitude correction is essentially negligible and the longitude correction makes adjustments on the order of 1--2 pixels. The correction becomes much more important for larger temporal spacing between observations.

Despite these challenges, using a number of different approaches we were able to obtain reasonably consistent estimates of the per pixel visual magnitude of the dust trail. Using a fully-calibrated L2 image from 2018 November 06 10:51~UT, we obtained the brightness values from 63 single-pixel locations centered along the visually brightest portion of the trail. We then derived an equivalent `background' for each pixel, where each value was the mean of four values obtained from the same physical sky location, absent the dust trail, in two images spaced approximately one hour before and after the target observation. Subtracting these backgrounds, we found the mean brightness along this background-subtracted trail to be 8.2$\times$10$^{-15}$ $B_\sun$ (where $B_\sun$ is the mean solar brightness) which using the currently available photometric calibration, equates to a visual magnitude of 15.6 (\textit{via P.Hess, Private Communication, October 2019}) per pixel. To support this value, we conducted similar analyses on adjacent images, and then performed the same process on the L3 data. In all cases our analyses returned values consistent to within a half a magnitude, primarily in the range 15.5--16.1. Thus we are comfortable to state a typical trail surface brightness of magnitude 15.8$\pm$0.3 per pixel.

While not analyzed in detail, a similar process applied to observations from the second {\it PSP} encounter returned values of the same order, validating the result we obtain in the first encounter. The observing conditions of the second encounter were largely the same as the first encounter, with both of these data sets calibrated similarly and having the same levels of uncertainty. Thus while the observations between successive encounters are self-consistent, we cannot rule out future modifications to the calibration factors that will slightly adjust the results presented here. That said, to validate our methodology we performed basic aperture photometry on a random selection of stars close to, but not overlapping, the dust trail in both encounters and found values to within $\pm{0.5}$ magnitudes of their Hipparcos catalog value, without accounting for different colors of stars. This reinforces our belief that our result for the dust trail brightness should be accurate to at least half a magnitude, though probably much better than this.

Our per pixel visual magnitude of 15.8, coupled with the 70$\arcsec$ per pixel resolution of WISPR-I, yields an estimated surface brightness of 25.0~mag~arcsec$^{-2}$. Although this is brighter than the limiting surface brightness of non-detections during Phaethon's December 2017 close approach (e.g., 27.2~mag~arcsec$^{-2}$ by \citealt{Jewitt2018}), the WISPR observations were made at much smaller solar elongations than can be achieved from most observatories. In $\S$\ref{s:discussion} we discuss these values in the context of non-detections and observability of the trail from Earth.

\subsection{Dust Properties\label{s:dust-properties}}
The faintness of the WISPR detection, coupled with the large pixel sizes of the WISPR instrument, limits the interpretations we can make of the dust properties of the trail. However, our photometric result allows us to estimate the total mass of dust in the entire trail.
First, we calculate the dust cross section per pixel using \citet{Jewitt1991}'s reformulation of \citet{Russell1916}:
\begin{equation}
    C_d = \frac{2.25{\times}10^{22}{\pi}r_\mathrm{H}^2{\Delta}^2}{p_V{\phi}({\alpha})}10^{0.4(m_{\odot} - m_{phaethon})}
\end{equation}
where $C_{d}$ is the dust cross section in m$^2$, $r_\mathrm{H}$ is the heliocentric distance of the dust trail in au, $\Delta$ is the {\it PSP}-trail distance in au, $p_V$ the geometric albedo,  $\alpha$ is the phase angle, $m_{\odot}$ is the apparent $V$ magnitude of the Sun ($-$26.74) and $m_{phaethon}$ is the estimated $V$ magnitude of the dust trail in one pixel (15.8). We assume the dust has the same geometric albedo (0.1066; \citealt{tedesco04}) and phase function (${\phi}(\alpha)$, which uses the {\it HG} parameters for Phaethon given in \citet{tabeshian19}) as Phaethon. The equation was evaluated for typical geometric parameters during the observations ($r_\mathrm{H}=0.15$~au, ${\Delta}=0.16$~au, ${\alpha}=52^\circ$). This yields $C_d = {\sim}4.1{\times}10^4$~m$^2$ per pixel near perihelion, where one pixel is 8100~km$\times$8100~km at Phaethon.

To estimate the total cross section of dust in the orbit, we must account for the lower spatial density of the dust near perihelion than elsewhere in the orbit. That is, if we assume the dust is uniformly distributed around the orbit, there should be equal amounts of dust per degree of mean anomaly. However, one degree of mean anomaly spans a longer linear distance near perihelion due to the higher velocity experienced at perihelion. We correct for this by scaling $C_d$ by the ratio of the velocity at perihelion ($v_{q}=106$~km~s$^{-1}$) to the average velocity around the orbit ($\bar{v}=20$~km~s$^{-1}$). This yields the average cross section of dust in a typical 8100~km~$\times$~8100~km region of the dust trail. We estimate that the trail is 14 WISPR pixels wide (113,400~km) all the way around the length (6~au) of Phaethon's orbit (in reality, the trail is wider far from perihelion, but this does not affect the amount of material), and thus find the total cross section of dust to be $3.4{\times}10^{11}$~m$^2$.
Finally, we assume that the trail is made of uniformly sized particles so that the total mass can be estimated as:
\begin{equation}
    m_{d} = \frac{4}{3}{{\rho}_d}{C_d}\bar{a}
\end{equation}
where ${\rho} = 2900$~kg~m$^{-3}$ is the bulk density of grains in the trail \citep{Babadzhanov2009} and $\bar{a} = 0.5$~mm is the average grain radius. This radius is of order the smallest grains that can be in a dust trail \citep{Sykes1992}. Its use assumes the optically-derived cross sectional area is dominated by such `small' dust. This yields a total mass estimate for the trail of $8.4{\times}10^{11}$~kg. Given our estimated photometric uncertainty of 0.3~mag, and an additional uncertainty of $\pm{4}$-pixels in trail width, this implies a range of total mass of $\sim(0.4-1.3){\times}10^{12}$~kg. This is lower than published estimates of the total mass of the Geminids from radar and visual observations ($10^{12}-10^{15}$~kg; \citealt{Ryabova2017}), but could be brought more in line by altering a variety of assumptions. However, this mass estimate is entirely inconsistent with a value based purely upon the recent production of dust by Phaethon. \cite{Jewitt2010} find perihelion activity to produce $\sim2.5{\times}10^{8}a_{mm}$~kg, where $a_{mm}$ is the dust radius in millimeters. Assuming our 0.5~mm radius particles, this yields an estimate of $\sim1{\times}10^{8}$~kg for the mass of a stream produced solely by perihelion activity on a single orbit. Even noting the estimated 250 years required for Phaethon to fill its orbit \citep{Ye2018}, the resulting total mass of $\sim1{\times}10^{10}$ kg is still much smaller than our estimated mass, implying that we are likely observing `old' dust (Geminids) rather than newer ($\leq$250 yr old) dust.

\section{Discussion} \label{s:discussion}

\subsection{Observations, and Relation to the Geminids}

Isolating the photometric signal from Phaethon's trail alone was challenging, with pixels encompassing the Phaethon trail having statistically the same photometric properties as bordering pixels. The trail location was visually identified in the L3 data, and verified to perfectly follow the Phaethon orbit, but for quantitative analyses in the L2 data we relied on the ephemeris predictions of the Phaethon orbit to extract data values. Despite these challenging observations, we are confident of our photometric estimate, with some caveats. Visually, we can clearly see that the signal is at the instrument noise level; this is most evident when trying to visually inspect small regions of the trail, when it becomes indistinguishable from the background. Thus, noise-level photometry was anticipated. The photometric values themselves are derived from a star-based calibration procedure similar to those successfully applied to the previous coronagraph and heliospheric imager instruments \citep{Bewsher2010, Colaninno2015}. A detailed description of WISPR calibration will appear in a publication currently under preparation \textit{(P.~Hess, Priv. Comm., October 2019)} and, while this study used the most up-to-date calibrations obtained directly from the WISPR team, there may be lower-level uncertainties and calibration issues that are currently not well defined. This is something that we can only address once we receive data from future encounters (which will occur approximately four times per year through 2024). Also, we restate briefly that we have not analyzed the portion of the trail that crossed the WISPR-O field of view due to non-linearities present in the photometry of those observations that remain under investigation by the WISPR team at the time of writing (\textit{P.~Hess, Priv. Comm., October 2019}).

Due to the very low signal intensity, inferences regarding the trail structure and density can not be made with any high degree of confidence. We do visually observe an apparent photometric gradient along the trail, with it being brightest near the right-hand edge of WISPR-I (e.g. Figure~\ref{fig:wispr-threepanel}) and becoming fainter and then undetectable at approximately the mid-point of the images. Loss of detection in this part of the field of view is likely be a result of the bright coronal streamers overwhelming the signal, but may also partially relate to increased photon noise nearer the Sun. However, we can not rule out the possibility that the trail density falls as we move from the perihelion point (see Figure~\ref{fig:wispr-threepanel}) and thus the cross section of dust per pixel falls below some critical threshold.

Related to this, we are yet to fully understand the variability of trail visibility throughout the encounter. Specifically, it is curious that we are barely able to visually detect the trail when it is at seemingly favorable phase angles, and at its closest point to the spacecraft, around 2018 November 10, and arguably not able to detect it at all for some of the time around 2018 November 08, although in the latter the trail is close to the slightly noisier field of view edge. This photometric behavior may hint at a complex scattering function, perhaps a consequence a broad range of particle sizes distributed throughout the trail. There may also be structural factors to consider as the gross structure of the Geminid stream is poorly understood, with recent studies finding it may be at least partially bifurcated \citep{Szalay2018}, for example.

\begin{figure}[ht!]
\centering
\includegraphics[scale=0.42]{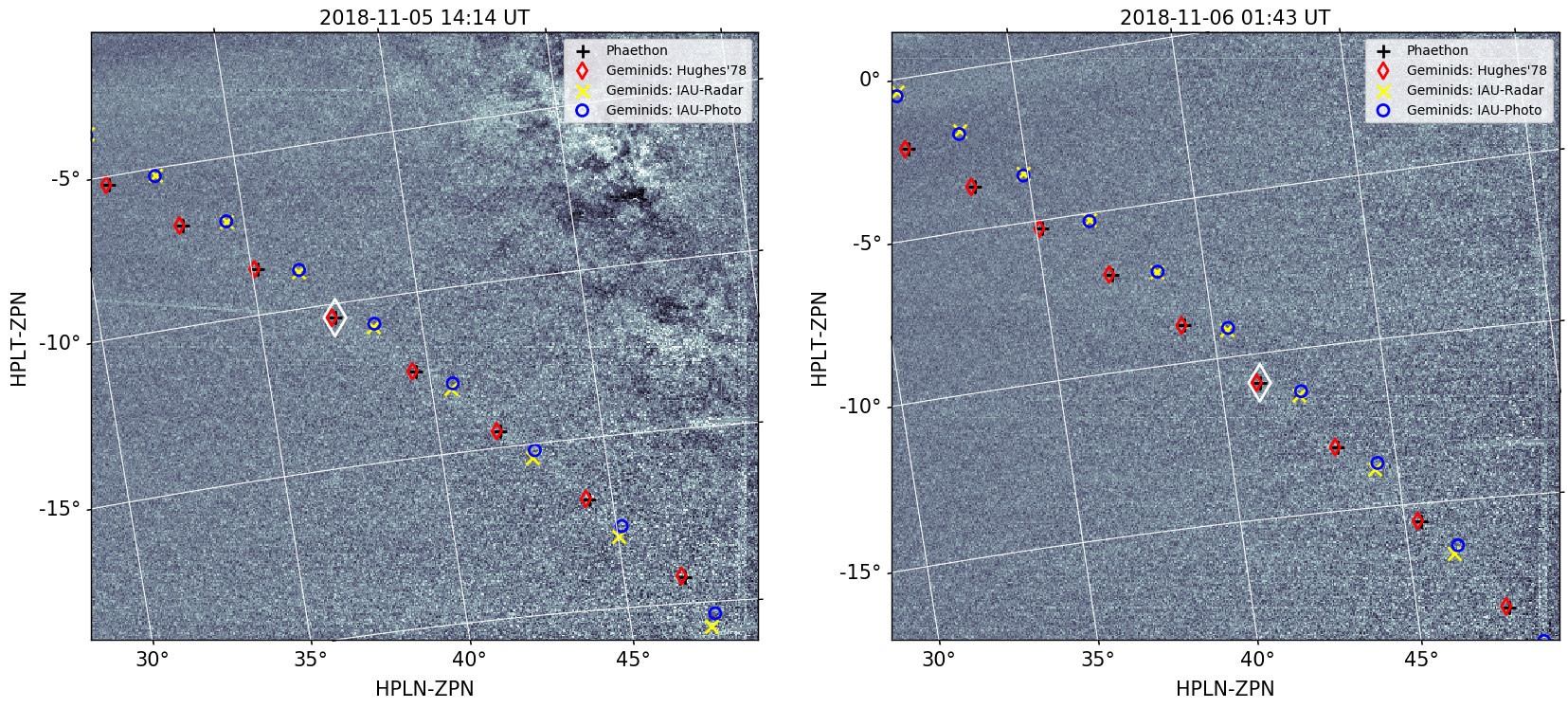}

\caption{Two WISPR observations, recorded on 2018 November 05 14:14~UT and 2018 November 06 1:43~UT, cropped so that the visible portion of the trail crosses the entire image. Black plus signs indicate the path of Phaethon through this image, and red diamonds, blue circles and yellow crosses indicating the path of Geminid orbits obtained from \cite{Williams1993} and integrated from their epoch of observation (1983) to November 2018. The large white diamond indicates the perihelion point of Phaethon, with pre- and post-perihelion as indicated in Figures~\ref{fig:wispr-threepanel} and \ref{fig:trail-evolution}.
\label{fig:geminid-orbits}}
\end{figure}

We can, however, make some general inferences about the structure of the Geminids by examining published orbits for individual meteors/streams. Figure 22.15 of \cite{Jenniskens2006} visualizes a range of orbits for different Geminid streams, showing the streams differing substantially in location near aphelion but naturally clustering around their perihelion. Thus we consider a hypothesis that WISPR is observing the clustering of the different Geminid streams at perihelion. We obtained three different values for Geminid orbits presented in \cite{Williams1993} - the Hughes orbit model \citep{Hughes1978}, and orbits derived from IAU Radar and Photographic data \citep{Williams1993} - and integrated them from their 1983 epoch to a November 2018 epoch, plotting the locations of the trails across the WISPR-I field of view on 2018 November 05 14:14~UT and 2018 November 06 1:43~UT. Figure~\ref{fig:geminid-orbits} shows portions of the WISPR data for those days, cropped to show just the part where the dust trail is most visible. As we can see, all four trails converge and overlap around perihelion, seemingly corresponding to the visually brightest portion of the dust trail. Outside of perihelion, two of the Geminid trails (IAU Radar and IAU Photographic trails) diverge quite substantially from the visible trail, but the Hughes trail continues to follow the apparent track of Phaethon and the dust trail throughout.

\begin{figure}[ht!]
\centering
\includegraphics[scale=0.36]{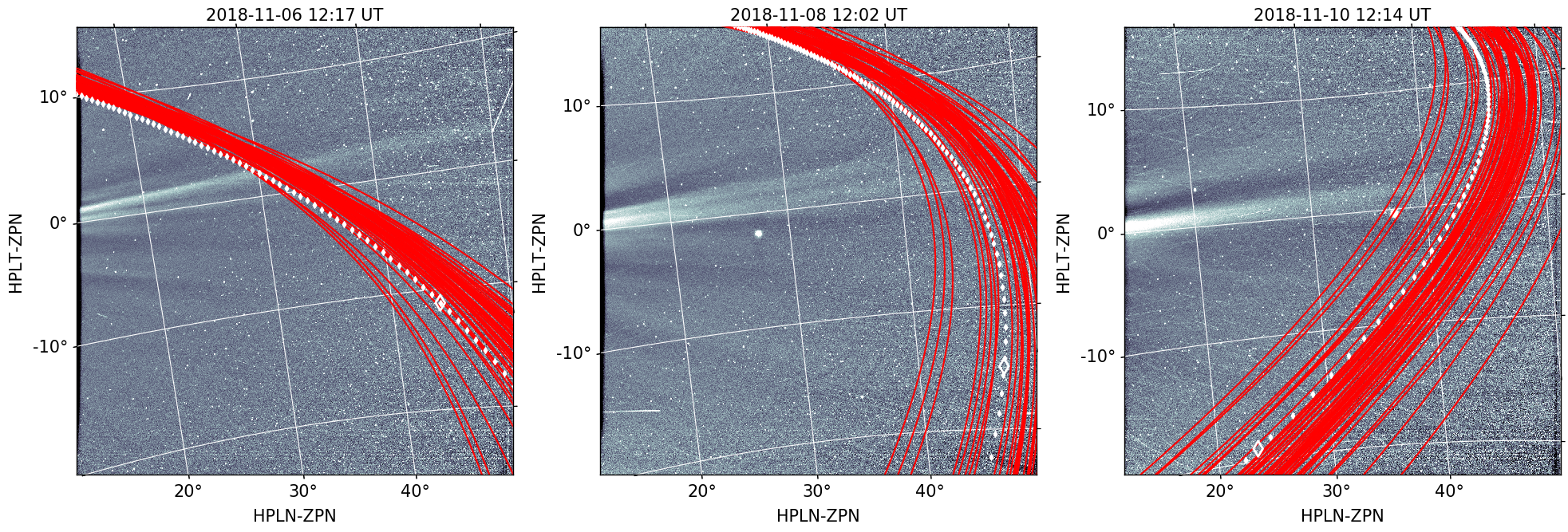}

\caption{Visualization of the orbits of 51 Geminids observed 2011 December 13--15 by the Canary Island Long-Baseline Observatory \citep{Koschny2013}. Superimposed (white) is the orbital path of Phaethon and (white diamond) its perihelion point.
\label{fig:geminid-orbit-evolution}}
\end{figure}

To further emphasize this point, a second example of Geminid orbits is shown in Figure~\ref{fig:geminid-orbit-evolution}, which presents a visualization of orbits calculated for 51 Geminid meteors observed 2011 December 13--15 \citep{Koschny2013}, again with white dots and a white diamond to represent the orbit of Phaethon and its perihelion point, respectively. We see that particularly early in the \textit{PSP} encounter (2018 November 5--6), the distribution of these individual objects is far more concentrated than later (2018 November 10, third panel) when we see a much wider distribution due to WISPR's changing viewing angle. Of course, caution must be taken to not over-interpret the trails shown in both Figures~\ref{fig:geminid-orbits} and \ref{fig:geminid-orbit-evolution} as these are essentially just samples of what is a broad distribution of objects (Geminids), some of which may never intersect with Earth, in a trail whose structure is undefined. However, we can use Figure~\ref{fig:geminid-orbit-evolution} in particular to mentally visualize a Geminid distribution that is likely centered close to both the orbit of Phaethon and the trail we observe in WISPR. Finally, we will reiterate here that the dust trail we observe remains perfectly centered on the Phaethon orbit throughout the encounter. The implications of this with regards to the Geminids are unclear, though imply that we are perhaps observing the largest mass particles that have remained closest to their original orbit at their original location of release (i.e. perihelion).
\newline
\subsection{Photometric Properties and Survey Non-detections}

Optical searches for a dust trail in Phaethon's orbit have, until now, been unsuccessful \citep[e.g.,][]{Ye2018, Jewitt2018}. Our per pixel visual magnitude estimate allows us to address the question of why WISPR is the first to detect this white-light trail or, alternatively, why the trail has not been seen in dedicated surveys, or in existing heliophysics imaging instruments.

We determined a per pixel visual magnitude of $\sim15.8$, corresponding to a surface brightness of 25.0~mag~arcsec$^{-2}$ as seen from \textit{PSP}'s location. This is well below the limiting surface brightness attainable by {\it SOHO} (limiting magnitude of $\sim$8.5 in C3 where pixels are 56$\arcsec$ yields 8.5 + $2.5{\times}$log$(56^2)$ = 17.2~mag~arcsec$^{-2}$) and {\it STEREO} (limiting magnitude of $\sim$13 in HI1 where pixels are 70$\arcsec$ yields 13 + $2.5{\times}$log$(70^2)$ = 22.3~mag~arcsec$^{-2}$). Thus, it is unsurprising that this trail has never been detected near its perihelion by these observatories, despite the latter detecting dust production by Phaethon at perihelion \citep[e.g.][]{Jewitt2012, Hui2017}.

When this trail is at solar elongations accessible to traditional telescopes, a number of factors reduce its surface brightness significantly lower. These factors are primarily the heliocentric distance (${\propto}r_\mathrm{H}^{-2}$), the spreading of the trail (described below), and the orbital speed of the dust (clustering)\footnote{We make no explicit correction for $\Delta$, because surface brightness is distance independent, but note that the surface brightness will actually increase as the trail's distance from the Sun increases due to the increasing linear density as discussed above.}. The latter effect will actually improve visibility as the local density of dust is higher as it moves slower in orbit, but is offset entirely by the former two effects, which reduce the surface brightness to a far larger degree. We can estimate the effect of each of these factors as follows.

First, regarding $r_\mathrm{H}$, we would expect the trail brightness to be fainter by $5{\times}$log(1~au / 0.15~au)$ = $+4.1~mag~arcsec$^{-2}$ at 1~au, and +6.0~mag~arcsec$^{-2}$ at aphelion (2.4~au). Second, regarding orbital spreading, Figures~\ref{fig:geminid-orbits} and \ref{fig:geminid-orbit-evolution} demonstrate that the Geminids do not maintain a uniform density throughout their orbit. They are largely focused at perihelion with just 0.00633~au separation between the Hughes and IAU Radar trails, for example, but spread significantly at aphelion with a separation of almost 0.2~au of those particular streams (which should be broadly reflective of the net behavior of the Geminid stream). This is a factor of $\sim$32 at aphelion and $\sim$16 at 1~au, which we can assume to represent a one-dimensional change in the trail surface area (that is, the trail spreads perpendicular to the orbit). Thus the surface brightness under these circumstances would be fainter by $2.5{\times}$log(16)$ = $+3.0~mag~arcsec$^{-2}$ (1~au) and $2.5{\times}$log(32)$ = $+3.8~mag~arcsec$^{-2}$ (2.4~au). Finally, regarding the clustering of the dust due to changing velocity, we find that surface brightness should be brighter by  $2.5{\times}$log[$v(r_\mathrm{H})/v(r_\mathrm{H}=q)]$ where $v(r_\mathrm{H})\propto\sqrt{2/r_\mathrm{H} - 1/a}$, $r_\mathrm{H}$ is heliocentric distance, $q$ is the perihelion distance, and $a$ is the semi-major axis of the orbit (1.27~au). At 1~au this yields a net brightening by $-$1.3~mag~arcsec$^{-2}$, and $-$3.1~mag~arcsec$^{-2}$ at 2.4~au. These values, and their net effect, are summarized in Table~\ref{tab:surface-brightness}. These corrections imply an equivalent surface brightness of 30.8~mag~arcsec$^{-2}$ at 1~au, which is considerably fainter than was achieved during the December 2017 close approach (27.2~mag~arcsec$^{-2}$; \citealt{Jewitt2018}), and demonstrates that the WISPR observations are effectively the deepest search ever for Phaethon's dust trail.

\begin{table}[ht!]
\centering
\begin{tabular}{ |p{5cm}||p{2.2cm}|p{2.2cm}|  }
    \hline
      & At 1~au & At 2.4~au\\
    \hline
    Heliocentric distance correction & +4.1 & +6.0 \\
    Orbital spreading & +3.0 & +3.8 \\
    Orbital velocity (clustering) & $-$1.3 & $-$3.1 \\
    \hline
    Net Effect & +5.8 & +6.7 \\
    Estimated trail brightness & 30.8 & 31.7 \\
    \hline
\end{tabular}
\caption{Estimate of factors affecting surface brightness of the Phaethon dust trail relative to {\it PSP} observations at perihelion. All values have units mag~arcsec$^{-2}$. \label{tab:surface-brightness}}
\end{table}

\subsection{Dust Properties}

It is apparent from our observations that the structure we observe fills the entire orbit. The observations presented here were recorded with Phaethon just three weeks beyond its aphelion, while we are observing the perihelion portion of the orbit. While not presented in detail here, observations of the same region of space recorded by WISPR during {\it PSP}'s second encounter (2019 March 30  - April 10) show a visually and photometrically identical dust trail, with Phaethon just three months from perihelion at that time. The implication from these observations is that the orbit is uniformly filled, a result consistent with the {\it COBE} DIRBE trail detection \citep{Arendt2014} which rather coincidentally occurred while Phaethon was near aphelion. \cite{Ye2018} argue that the Phaethon orbit fills in $\sim$250 years, so it is perhaps not surprising that we can confirm this, and the Geminids are known to encompass their entire orbit.

We find a total mass of dust in the presumed Phaethon orbit to be significantly more than it is possible for Phaethon to produce at its current levels \citep{Jewitt2012}, but somewhat less than previously published estimates for the mass of the Geminids \citep{Ryabova2017}, though the latter study does state a preference towards the lower end of their 10$^{12}$--10$^{15}$~kg mass estimate. It is easier to reconcile our observations as being a large underestimate of the mass of the Geminids as opposed to a large overestimate of the dust produced at perihelion by Phaethon, and thus we can assume that the trail we observe is primarily Geminids. It is impossible to rule out that our signal contains some contribution from recently produced dust by Phaethon, but this would represent a very small fraction of the stream mass we calculate.

Broadly speaking, we believe our low mass estimate to be a consequence of the fact that we can barely detect even the brightest part of the stream. However, to bring our mass estimate in line with the assumed mass of the Geminids would require the trail brightness to be of the order (V) magnitude 12 per pixel in the observations (using the same assumptions). Not only do we see no evidence for such an error on the WISPR calibration, if the trail brightness was of the order visual magnitude 12 per pixel (or even 13--14) its surface brightness would be within the reach of the \textit{STEREO} HI-1 instrument, which does not detect the trail despite observing the perihelion portion of the orbit, as well as Phaethon itself. A number of other assumptions within our dust mass calculation could also lead to an underestimate of the total mass. One consideration is that we have assumed grains of similar size in the trail but, as noted in \cite{Jewitt2013}, impact flashes from Geminid meteors have implied masses of up to 5~kg \citep{Yanagisawa2008}. Just a factor of ten adjustment to our assumed average grain size would push our mass estimates more comfortably within the lower bounds of published values. In reality, the dust grain sizes likely follow a power-law and, depending on the choice of power-law, the majority of the visible flux is usually dominated by the choice of the smallest particles while the mass is dominated by the choice of the largest particles. Some constraints on the mass index of the Geminid stream have been provided by both visual and radar observations, in yielding indices of around 1.7 \citep[e.g.][]{Arlt2006, Blaauw2011}, with the most recent comprehensive study using optical, radar and lunar impact observations to determine a mass index of 1.68$\pm$0.04, with mass bounds of $10^{-6}-10^{3}$~g \citep{Blaauw2017}. The implication is that the Geminid mass is dominated by heavier particles, reinforcing our belief that we are underestimating the mass of the stream. Finally, we have shown that we likely are observing a clustering of just some portion of the Geminid stream following closely to the perihelion portion of their assumed parent's orbit. This conclusion, if correct, further complicates our dust mass estimations as we are likely only sampling a subset of the entire stream distribution, and likely not a portion of the stream that is ever encountered (sampled) by Earth. Such details are unconstrained by the WISPR data and, rather than speculating further, we simply note that our mass estimate for this dust trail appears to be in plausible agreement with estimates of the total mass of the Geminids.

\section{Conclusions} \label{s:conclusions}

We have presented the first detection of a white-light dust trail following the orbit of asteroid (3200) Phaethon, detected by the WISPR instrument in the first two encounters of the NASA \textit{PSP} mission. The key results of this study are:

\begin{enumerate}
    \item Visual identification of a white-light dust trail appearing to follow the perihelion portion of Phaethon's orbit.
    \item Determination of the visual (V) magnitude of the stream to be 15.8~$\pm$~0.3 per pixel and a surface brightness of 25.0 mag arcsec$^{-2}$.
    \item An estimate of the total mass of the stream to be $\sim(0.4-1.3){\times}10^{12}$~kg, plausibly consistent with (though below) the assumed mass of the Geminid stream but greater than could be produced by activity of Phaethon at perihelion.
\end{enumerate}

Our visual magnitude and corresponding surface brightness calculations demonstrate why this trail has not be detected at optical wavelengths to date, despite dedicated surveys. We are able to provide a number of reasons why our result for the estimated mass of the entire trail would be an underestimate of the true value, supporting a conclusion that we are likely seeing a clustering of some part of the Geminid stream near perihelion as opposed to more recently ($\leq$250 yrs) released dust from perihelion activity of Phaethon.

We are fortunate that the planned orbit for \textit{PSP} brings the spacecraft to perihelion in approximately the same location in every future encounter for the next several years. Thus we will have the unique opportunity to study Phaethon's dust trail approximately every three months for the mission duration. In addition to performing the presented analyses on successive encounters, we plan to also look at temporal trends in the brightness of the stream as a function of Phaethon's location in its orbit. Once we have obtained additional observations, and the photometric properties of the WISPR instrument are fully characterized, we also plan to undertake a more detailed dust-modeling study to better characterize the role of phase angle and viewing geometry on the visibility of the trail. These observations and results may be of critical value to the Japan Aerospace Exploration Agency (JAXA) proposed Demonstration and Experiment of Space Technology for INterplanetary voYage (DESTINY$^+$) \citep{Arai2018} mission, which would launch in 2022 and perform a flyby of Phaethon during its planned $\sim$4 year mission to study the asteroid and its dust environment \citep{Kruger2019}.

\acknowledgments

K.B., R.A.H., G.S. and B.M.G. were supported by the NASA/WISPR. M.M.K., and M.S.P.K. were supported by NASA Near Earth Object Observations grant No. NNX17AK15G. The authors would like to acknowledge P. Hess (NRL) for his help with the calibration of the WISPR observations. Parker Solar Probe was designed, built, and is now operated by the Johns Hopkins Applied Physics Laboratory as part of NASA's Living with a Star (LWS) program (contract NNN06AA01C). Support from the LWS management and technical team has played a critical role in the success of the {\it Parker Solar Probe} mission. The Wide-Field Imager for Parker Solar Probe (WISPR) instrument was designed, built, and now operated by the US Naval Research Laboratory (contract NNG11EK11I) in collaboration with Johns Hopkins University/Applied Physics Laboratory, California Institute of Technology/Jet Propulsion Laboratory, University of Gottingen, Germany, Centre Spatiale de Liege, Belgium and University of Toulouse/Research Institute in Astrophysics and Planetology. We are grateful to an anonymous referee, whose comments have substantially improved the presentation of our results.


\vspace{5mm}


\software{AstroPy \citep{Astropy13},
          NumPy \citep{Numpy11},
          MatPlotLib \citep{Matplotlib07}
          }

\bibliography{phaethon_wispr}

\end{document}